\documentclass[10pt,journal]{IEEEtran}

\usepackage[utf8]{inputenc}
\usepackage{graphics,graphicx,color}

\usepackage[table]{xcolor}
\usepackage{xcolor, color, soul}
\usepackage{amsmath,amssymb,amsthm,amsfonts,bm,bbm}
\usepackage{algorithmicx,algorithm,algpseudocode}
\usepackage{multirow}
\usepackage{cite}
\usepackage{url}
\usepackage[normalem]{ulem}
\usepackage[english]{babel}
\usepackage{graphicx}
\usepackage{framed}
\usepackage{multirow}
\usepackage[normalem]{ulem}
\usepackage{amsmath}
\usepackage{amsthm}
\usepackage{amssymb}
\usepackage{amsfonts}
\usepackage{enumerate}
\usepackage[utf8]{inputenc}
\usepackage{caption}
\usepackage{float}
\usepackage{hyperref}
\usepackage{tikz}
\usepackage{authblk}
\usetikzlibrary{shapes.geometric, arrows}
\usetikzlibrary{automata, positioning, shapes.multipart,fit}
\tikzstyle{data} = [rectangle,  minimum width=1cm, minimum height=0.75cm,text centered, draw=gray, fill=gray!30]
\tikzstyle{process} = [circle, minimum size =1cm, text centered, draw=gray, fill=gray!30]
\tikzstyle{process_h} = [circle, minimum size =1cm, text centered, draw=blue, fill=blue!30]
\tikzstyle{arrow} = [thin,->,>=stealth]

\sethlcolor{lightgray}
\sethlcolor{yellow}

\newcommand{\sgn}{\operatorname{sign}}
\newcommand{\grad}{\operatorname{grad}}
\newcommand{\tr}{\operatorname{tr}}
\DeclareMathOperator*{\argmax}{\arg\max}
\usepackage{subcaption}
\usepackage{todonotes}
\presetkeys{todonotes}{color=blue!20}{}

\usepackage{color}
\definecolor{yellow}{rgb}{1,1,0.7}

\usepackage{footmisc}
\usepackage[normalem]{ulem}
\usepackage{comment}
\title{Binarized ResNet: Enabling Robust Automatic Modulation Classification at the resource-constrained Edge}

\author{Deepsayan Sadhukhan*, Nitin Priyadarshini Shankar*, Nancy Nayak, Thulasi Tholeti and Sheetal Kalyani\\

\thanks{\hspace{-0.7cm}\textbf{*}Equal contribution. \\
The authors are with the Department of Electrical Engineering, Indian Institute of Technology Madras, India. Emails: \{ee20s001@smail, ee20d425@smail, ee17d408@smail, ee15d410@ee,
skalyani@ee\}.iitm.ac.in}
}

\begin{document}
	\maketitle
 \ifCLASSOPTIONonecolumn
\vspace{-1.5cm}
\fi
	\begin{abstract}
		Recently, deep neural networks (DNNs) have been used extensively for automatic modulation classification (AMC). Typically, DNNs are unsuitable for deployment at resource-constrained edge networks due to their high complexity. They are also vulnerable to adversarial attacks, which is a significant security concern. This work proposes a rotated binary large ResNet (RBLResNet) for AMC that can be deployed at the edge network because of low complexity. The performance gap between the RBLResNet and existing architectures with floating-point weights and activations can be closed by two proposed ensemble methods: (i) multilevel classification (MC), and (ii) bagging multiple RBLResNets. The MC method achieves an accuracy of $93.39\%$ at $10$dB over all the $24$ modulation classes of the Deepsig dataset. This performance is comparable to state-of-the-art performances, with $4.75$ times lower memory and $1214$ times lower computation. Furthermore, RBLResNet also has high adversarial robustness compared to existing DNN models. The proposed MC method with RBLResNets has an adversarial accuracy of $87.25\%$ over a wide range of SNRs, surpassing the robustness of existing methods to the best of our knowledge. Low memory, low computation, and the highest adversarial robustness make it a better choice for robust AMC in low-power edge devices.

	\end{abstract}
	
    \section{Introduction}
Automatic modulation classification (AMC) has become a popular tool in recent years for identifying the modulation type of transmitted signals. It is predominantly employed in systems that use adaptive modulation and coding. It is necessary to classify the modulation type of the signal before performing any signal processing at the receiver; AMC must be performed in real-time to avoid control overhead. With the advent of multiple-input-multiple-output (MIMO) technology in beyond 5G massive machine type communication (mMTC), a huge number of receivers receive signals with different modulation schemes from a broader range of sources. When the receivers are resource constrained in terms of memory and computation, a memory-efficient AMC method of low complexity would be of practical use.

Traditional AMC methods based on likelihood \cite{huan1995likelihood, wei2000maximum, ramezani2013likelihood} achieve optimal solutions theoretically at the cost of high computational complexity. On the other hand, the feature-based AMC methods, consisting of a feature extractor (FE) and a classifier, reach a sub-optimal solution with very low complexity. Recently, deep learning (DL) based FEs \cite{meng2018automatic,o2016convolutional, huang2019automatic, hermawan2020cnn, peng2018modulation, mendis2017deep, rajendran2018deep, o2018over} have gained immense popularity for extracting the best combination of features automatically while adapting to the set of modulations and different channel environments.

In MCNet \cite{8963964}, a convolutional neural network (CNN) is used to learn the spatio-temporal correlations using asymmetric kernels and skip connections. However, adding more blocks to improve accuracy leads to an increase in complexity and memory requirement. In InvoResNet \cite{zhang2021automatic}, the authors proposed an involution method to enhance the discrimination capability and expressiveness of the model for faster convergence at the cost of huge computational complexity. An LSTM block appended at the end of residual blocks helps to achieve similar accuracy with much lesser complexity \cite{1810}. 

Various methods for AMC such as constellation-based feature extraction using DL \cite{182}, multi-task learning \cite{181}, cascaded deep neural networks (DNNs) for FE and classification \cite{184}, and Transformer-based AMC \cite{189} achieve good accuracy but at the cost of high computational complexity, and therefore, not useful to deploy in the edge network. A multiscale-CNN-based FE was proposed in \cite{188}, but it performs worse at lower SNRs. Another technique for AMC uses a combination of CNN and gated recurrent unit \cite{1811}, but it does not use the data from the complete range of SNRs. 

Typically, DL-based FEs are `power-hungry' and have an excessive signal processing requirement at the user end. Therefore, they are unsuitable for deployment at the receiver of mobile devices on the edge network. Recently, there has been an increased interest in FEs that are computationally less demanding in order to achieve lower latency without any degradation in performance \cite{wang2010fast}. To reduce the number of parameters, Lightweight\cite{kim2020lightweight} introduced asymmetric kernel dimensions. A distributed learning-based AMC is proposed in \cite{wang2020distributed} because of its lower computing overhead. 

One way to significantly reduce the memory and compute requirements is to design a binary neural network (BNN) suited to our application \cite{courbariaux2016binarized, rastegari2016xnor, he2020proxybnn}. For example, in \cite{9505671}, the authors propose a BNN-based neural Turbo Auto Encoder that can perform as well as existing methods in literature while saving in terms of memory and computational complexity. However, when applied to AMC applications, binarized versions of the existing SOTA architectures do not learn anything meaningful due to a drastic loss in representation capability, causing a huge quantization error. In this paper, inspired by the architectures used for the image classification problem \cite{he2016deep} and using domain knowledge related to the problem of AMC, we design a ResNet-based architecture with high representation power and name it LargeResNet or \textit{LResNet.} It is due to this high representation power that a proposed binary version of LResNet, named \textit{BLResNet}, is able to at least learn a model. However, the performance of BLResNet is still not adequate since there is a large performance gap compared to LResNet \cite{hubara2017quantized, rajendran2018deep}.
Even in the case of a binary network constructed with real first and last layers, there is no significant increase in accuracy. Hence, we attempt to first close this gap partially by designing a rotated BNN (RBNN) \cite{lin2020rotated}. 

RBNNs reduce the angular bias between the real-valued parameter and its binary version during training leading to reduced quantization error upon binarization. However, the RBNN version of any arbitrary architecture does not perform as well as its real\footnote{Architectures with floating point weights and activations are referred to as real architectures.} counterpart. The rotated binarized version of LResNet, named \textit{RBLResNet}, improves the performance of BLResNet to a great extent.
To improve the performance of RBLResNet even further while keeping the complexity low, we propose to ensemble multiple BNNs/RBNNs such that the resultant network has enhanced performance than each of the weak learners. We first propose bagging \cite{zhu2019binary}, where a weighted sum of the output probabilities of multiple weak learners (RBLResNet) is taken. Next, we propose multilevel classification (MC), where the more complex classification problem is broken down into simpler problems which are then solved by multiple instances of the same RBLResNet. The resultant models' performance is very close to SOTA solutions for AMC with very low memory and computation power.

AMC is adopted for spectrum monitoring and the analysis of intercepted signals in not only civilian applications but also military systems \cite{gouldieff2017blind,hoang2019automatic, kim2016deep, haring2010automatic, shi2001gabor}. Therefore, the security aspect is as crucial as reducing the memory and computational requirements. It has been shown that DNNs are highly vulnerable to adversarial samples \cite{szegedy2013intriguing}. Adversarial samples are malicious inputs constructed by perturbing the input data point by a minimal value. The perturbation is made in such a way that the model misclassifies an otherwise correctly classified data point. The works \cite{sadeghi2018adversarial,usama2019black} have shown how the DNNs for AMC can be attacked to misclassify even with a minimal transmitting power of the adversary. 

Adversarial training is used to achieve robustness against weak attacks in \cite{maroto2022safeamc} and \cite{sahay2021robust}; the latter also uses an autoencoder-decoder to detect high-intensity attacks. However, the computational cost of adversarial training and decoding grows prohibitively as the size of the model and the number of input dimensions increase. Further, training against less expensive and weaker adversaries produces robust models against weak attacks but breaks down under stronger attacks. The BNNs are inherently more adversarially robust than networks with full-precision weights and activations without explicit adversarial training \cite{galloway2018attacking}. Therefore, we investigate the robustness of the proposed RBLResNet and the ensemble methods like bagging and MC in the context of AMC. 

The key contributions of this work are,
\begin{itemize}
        \item Using the domain knowledge of AMC, we propose a ResNet-based architecture LResNet that has a high representation power and is suitable for binarization.
        \item We also propose a rotated binarized network RBLResNet for AMC. The performance of RBLResNet is as good as the SOTA architectures while having $64$ times lesser memory requirement and $64$ times more speed when compared to LResNet.
        \item We further propose two types of ensemble techniques with RBLResNet: (i) multilevel classification and (ii) bagging that improve the accuracy by $6.44\%$ and $2.75\%$, respectively. The MC method with RBLResNets achieves an accuracy of $93.39\%$, which is close to the performance of InvoResNet with a $4.75$ times lower memory and $1214$ times lower computation.
        \item We also show that rotated binarization improves robustness against white-box attacks\footnote{White-box attacks involve an attacker possessing comprehensive knowledge about the deployed model, including details like inputs, model structure, and other specific information.}. While Lightweight \cite{kim2020lightweight} has an adversarial accuracy of $59.72\%$, the proposed RBLResNet has an adversarial accuracy of $73.87\%$. Further, bagging and MC significantly improve the adversarial robustness producing an accuracy of $79.53\%$ and $87.25\%$, respectively. 
\end{itemize}
We describe the system model in Sec. \ref{sec:systemmodel}. In Sec. \ref{sec:proposedmethod}, we discuss the proposed methods in detail and then provide extensive simulations in Sec. \ref{sec:radioML}, which proves its aptness to use in edge networks.

\section{System Model}
\label{sec:systemmodel}
\begin{figure}[!t]
    \centering
   
    \hspace{3.7cm}\includegraphics[scale=0.6]{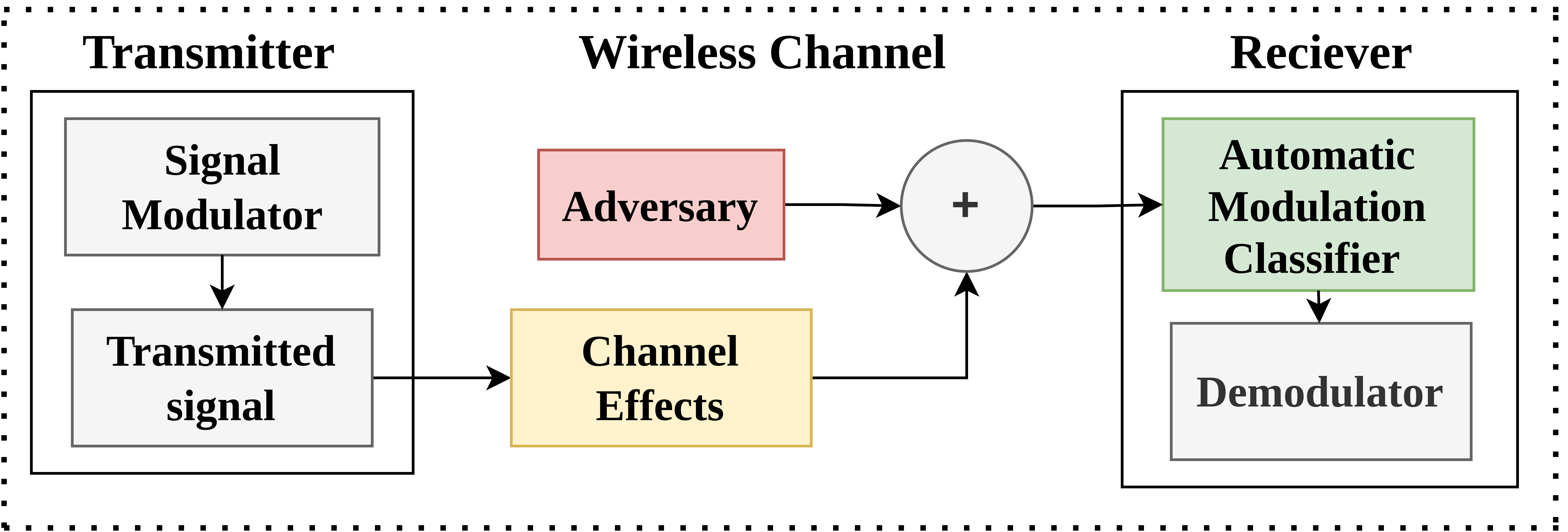}
    \caption{Wireless system model}
    \label{fig:SystemModel}
\end{figure}

A $k$-dimensional modulated signal having a constant normalized symbol rate across all modulation schemes is generated, denoted by $\mathbf{s}=\left[s[0],\,s[1],\,...,\,s[k-1]\right]^T$ and is transmitted over the air having various channel effects. The channel is considered to have the following effects - sample rate offset (SRO), center frequency offset (CFO), selective fading models (Rician and Rayleigh), and finally, the additive white Gaussian noise (AWGN). The $l$-th sample of the received signal $\mathbf{x}$ is given by
\begin{equation*}
    x[l]=s[l] \ast h[l] + n[l],
\end{equation*}
where $\mathbf{h}$ represents the channel, $*$ denotes the convolution operator, and the noise $\mathbf{n}$ added to the received signal is complex AWGN, with each sample distributed as $\mathcal{CN}(0, N_0)$. The in-phase (I) and quadrature (Q) components of the signal are used to represent all signals as two-dimensional reals, despite the fact that they are all complex. Fig. \ref{fig:SystemModel} represents the block diagram for the system model. Besides channel and noise effects, adversarial interferences can further degrade the transmitted signal. In an adversarial setting, the attacker adds an adversarial component denoted by $\mathbf{\zeta}$ to the clean signal and the received signal is,
\begin{equation*}
    x^{adv}[l]=x[l]+\zeta[l].
\end{equation*}

Based on the availability of model information to the attackers, there are two types of attacks, (i) white box and the (ii) black box. The white box attacks are those in which the adversary has complete access to the target system's model information, including weights and gradients. In black box attacks, the adversary can only access the outputs from the model. Compared to black-box attacks, white-box attacks are more precarious. Our work primarily concentrates on two types of white-box attacks discussed below.
\begin{itemize}
    \item \textbf{Fast Gradient Sign Method:} For Adversarial Attacks, we take the fast gradient sign method (FGSM), which is a well-known attack against DNNs \cite{goodfellow2014explaining}. The adversary of the input signal $\mathbf{x}$ is generated as follows: 
\begin{equation*}
    \mathbf{ x}^{adv} = \mathbf{x} + \epsilon\, \text{Sign}(\nabla_{\mathbf{x}} J(f(\mathbf{x}), y))
\end{equation*}
where $y$ is the corresponding label of $\mathbf{x}$. $J$ is the loss function with respect to the input sample for the function $f$. The gradient with respect to the input $\mathbf{x}$ is denoted by $\nabla_\mathbf{x}$. The intensity of the attack is characterized by the multiplier $\epsilon $, whose value is generally kept small between $0$ to $0.1$.

\item \textbf{Projected Gradient Descent Method: } As FGSM is a single-step attack, the input is perturbed once. projected gradient decent (PGD) \cite{madry2017towards} is an iterative attack where the adversarial sample is found in the same way as FGSM but iteratively starts from a random point on the norm-ball and updates the sample according to 
\begin{equation*}
         \mathbf{x}^{\text{adv}}_{i+1} = \text{Proj}_{B_{\xi} (\mathbf{x})}\left( \mathbf{x}^{\text{adv}}_{i} + \eta \, \text{Sign}\left( \nabla_{\mathbf{x}^{\text{adv}}_{i}}J(f(\mathbf{x}^{\text{adv}}_{i}),y)\right)\right),
\end{equation*}
where $\text{Proj}_{B_{\xi}( \mathbf{ x}^{adv})} (\mathbf{x}')= \arg \min_{\mathbf{ x}' \in B_{\xi}(\mathbf{ x})} \|{\mathbf{x}^{adv}-\mathbf{x}'}\|_p$ is how the perturbation distance is minimized iteratively. The adversary determines the parameters such as the number of iterations, $\xi$, and the step size $\eta$.

\end{itemize}

In the next section, we propose a ResNet-based architecture named LResNet followed by its rotated binarized version named RBLResNet for AMC. Further, we propose two ensemble techniques that allow the deployment of AMC at the resource-constrained edge.

\section{Proposed Method}
\label{sec:proposedmethod}

\begin{figure}[!t]
    \centering
    \begin{subfigure}{0.37\linewidth}
    \includegraphics[scale=0.6]{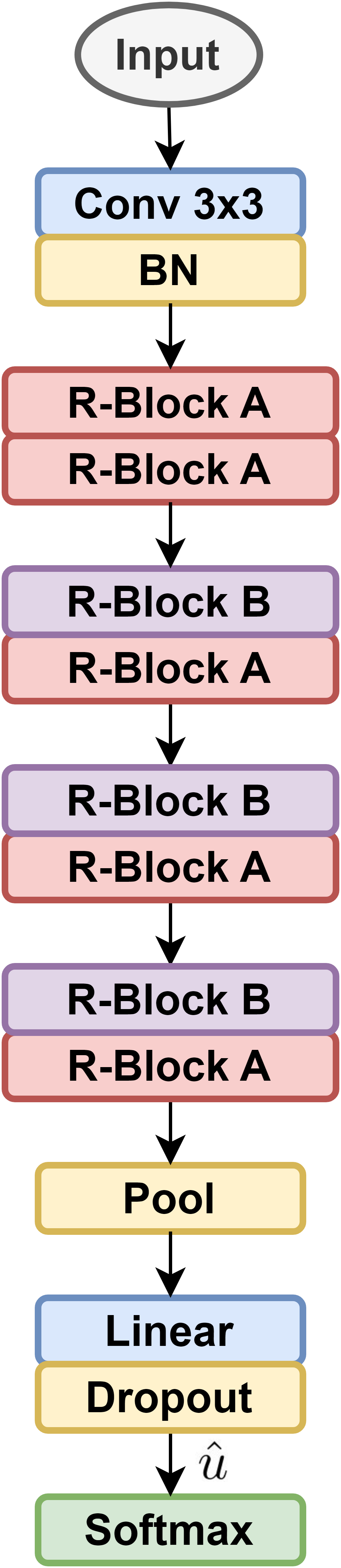}
    \caption{Overall Architecture}
    \end{subfigure}%
    \begin{subfigure}{0.25\linewidth}
    \includegraphics[scale=0.55]{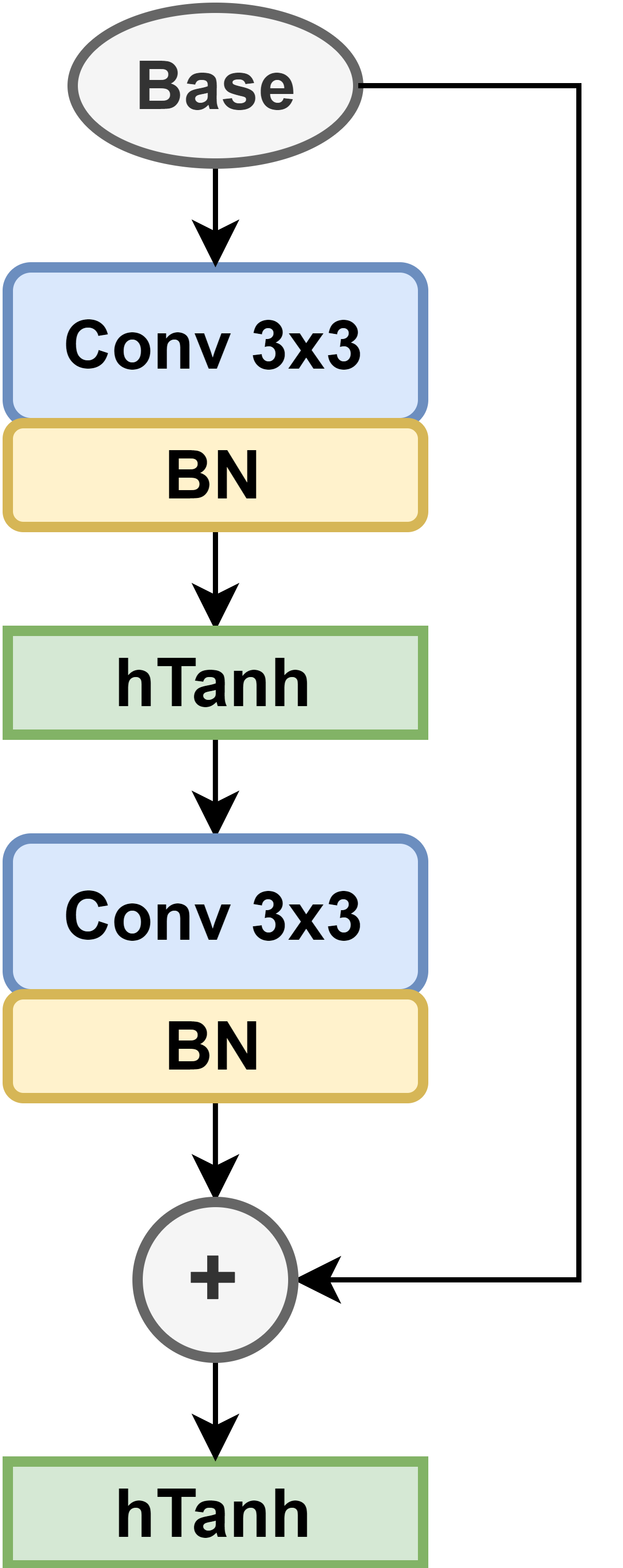}
    \caption{R-Block A}
    \label{fig:ResnetA}
    \end{subfigure}%
    \begin{subfigure}{0.25\linewidth}
    \includegraphics[scale=0.55]{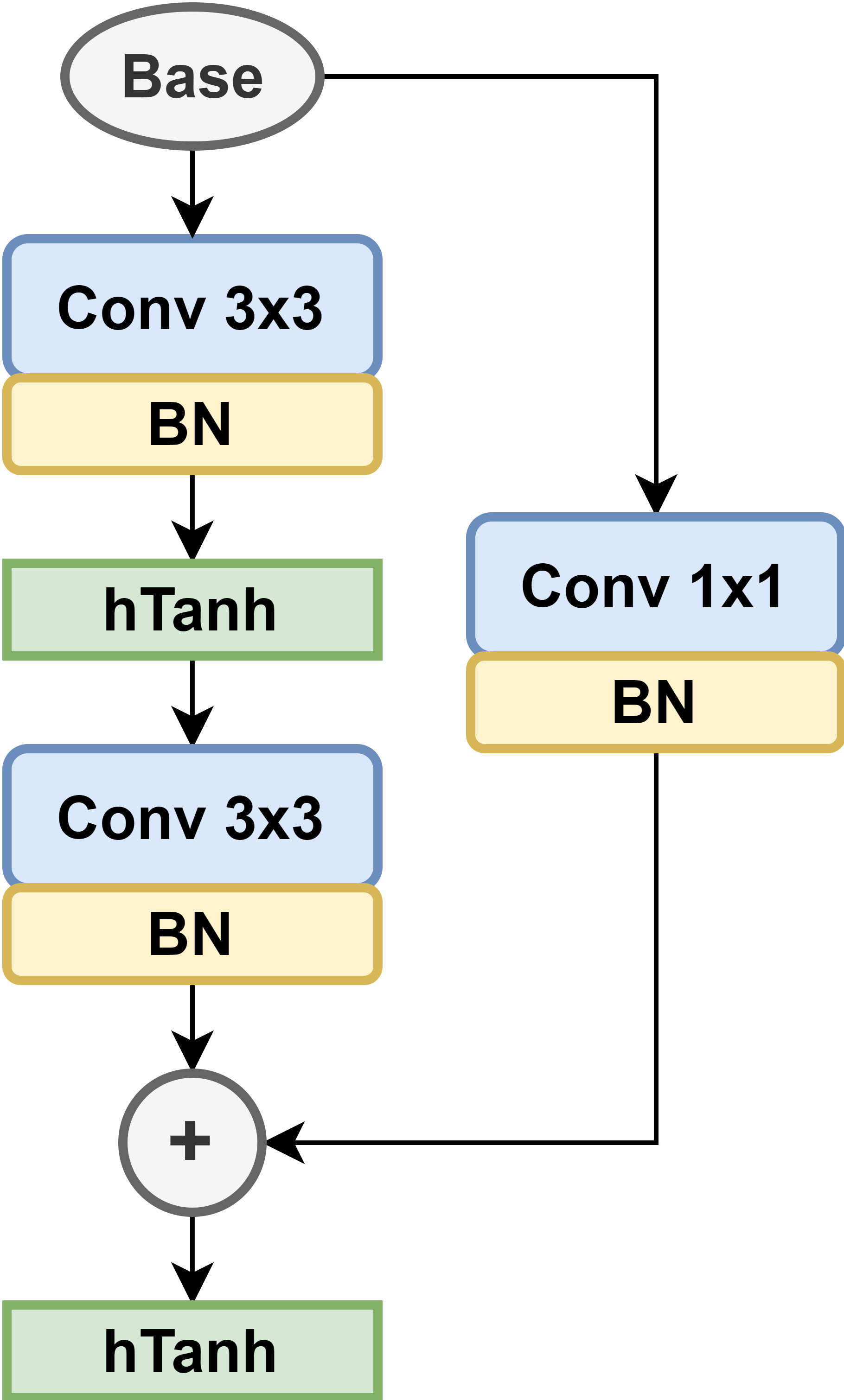}
    \caption{R-Block B}
    \label{fig:ResnetB}
    \end{subfigure}
    \caption{Proposed architecture} 
    \label{fig:Architecture}
\end{figure}

Inspired by the image classification problem \cite{he2016deep}, we propose a residual network architecture LResNet with a higher number of output filters with increasing depth that increases the number of parameters, leading to higher representation power, making it suitable for rotated binarization. The core of the proposed architecture contains Residual units whose couple is called a ResNet block. The main advantage of a ResNet block is the presence of skipped connections which adds the input to a later stage of the block enabling the network to learn the Residual. It helps the network to overcome the vanishing gradients problem. Each ResNet block has 2D convolutional layers as its fundamental entity, as shown in Fig. \ref{fig:Architecture}. The architecture mainly has two different types of Residual blocks, as illustrated in Fig. \ref{fig:ResnetA} and Fig. \ref{fig:ResnetB} according to whether we need to reduce the spatial dimension or not. The only difference between the two blocks is the addition of a convolution layer with $1\times1$ kernel size and stride $2$ to reduce the spatial dimension. On placing these blocks alternatively, the horizontal dimension is reduced gradually, decreasing the computation complexity of the network.

Instead of re-using an existing architecture suitable for the image classification task, we use the domain knowledge of AMC to customize the architecture for AMC and reduce its complexity. Unlike the input in image classification tasks that has three channels, `R', `G', and `B', the AMC problem has only one channel. Therefore, the first convolution layers of the proposed architectures have only a single input channel. The domain knowledge is also used to reduce the dimension of the filters. The initial layers of the proposed architecture contain symmetric 2D filters of dimension $(3 \times 3)$ that can extract the features omnidirectionally. This means the filters can find patterns between the I component at time $t$ and the Q component at time $t-1$. Suppose we consider using filters of size $1 \times 3$ or $3 \times 1$ at the initial layers of the architecture. In that case, we will miss out on these patterns because the filters will either be looking at the same instant of time or will be looking only at the I component or Q component and not both across different time instants. The first convolutional layer of the 1st R-Block B reduces the spatial dimension of the input from two to one and thereby eradicating the need for two-dimensional feature extraction ($3 \times 3$ filters) in the rest of the blocks and 1D filters of size $1\times3$ are used.

As we move further into the architecture, the number of filters gradually increases from $32$ to $128$ in a span of $4$ blocks extracting the finer details as we go in-depth. The omnidirectional feature extraction, along with the increased number of filters, ensures that the representation capability of the architecture is increased. The first convolutional layer is designed to extract general features without altering the spatial dimension of the input and has $32$ filters of shape $3\times3$. After that, a $2D$ batch normalisation (BN) is performed along the spatial dimensions of the output of the convolutional layer. It is practised for all the convolutional layers since it increases the stability and reduces the number of epochs required for training. The \textit{hard tanh} activation function is used in all the blocks except at the output, where a softmax activation is used. After stacking several blocks, we perform an average-pooling operation followed by a $2D$ BN and a linear layer (with softmax activation) to perform the classification. A dropout layer is included to prevent over-fitting. 

It can be seen in Table. \ref{tab:table2} that the FLOP count and memory requirements of the proposed LResNet are at the higher end while it outperforms most of the SOTA architectures with similar order of complexity. Given the current focus on green communication and low-cost edge devices, we propose to use smart binarization and ensemble techniques on the proposed real architecture in the subsequent subsections.

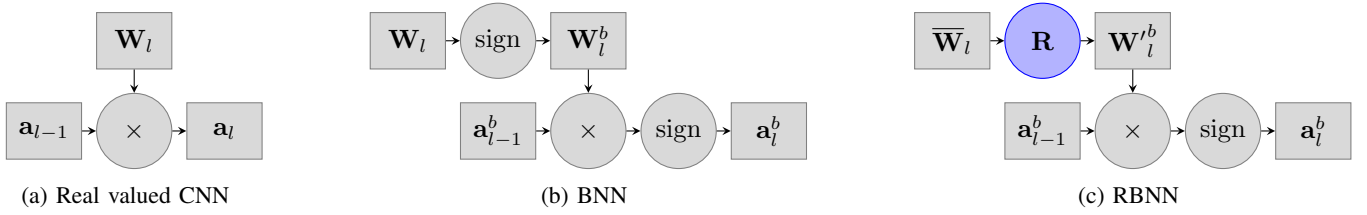
\begin{figure*}[h!]
   \subfloat[Real valued CNN \label{fig:CNN}]
    {
        \begin{tikzpicture}[node distance=1.2cm]
        \node (Wl+1) [data] {$\mathbf{W}_{l}$};
        \node (x) [process, below of=Wl+1] {$\times$};
        \node (al) [data, left of=x] {$\mathbf{a}_{l-1}$};
        \node (al+1) [data, right of=x] {$\mathbf{a}_{l}$};
        \draw [arrow] (Wl+1) -- (x);
        \draw [arrow] (al) -- (x);
        \draw [arrow] (x) -- (al+1);
        \end{tikzpicture}
    }
\hspace{\fill}
   \subfloat[BNN \label{fig:BNN} ]  
   {%
        \begin{tikzpicture}[node distance=1.2cm]
        \node (Wl+1) [data] {$\mathbf{W}_{l}$};
        \node (sign) [process, right of=Wl+1] {$\sgn$};
        \node (Wbl+1) [data, right of=sign] {$\mathbf{W}_{l}^b$};
        \node (x) [process, below of=Wbl+1] {$\times$};
        \node (abl) [data, left of=x] {$\mathbf{a}_{l-1}^b$};
        \node (sign2) [process, right of=x] {$\sgn$};
        \node (abl+1) [data, right of=sign2]{$\mathbf{a}_{l}^b$};
        \draw [arrow] (abl) -- (x);
        \draw [arrow] (Wl+1) -- (sign);
        \draw [arrow] (sign) -- (Wbl+1);
        \draw [arrow] (Wbl+1) -- (x);
        \draw [arrow] (x) -- (sign2);
        \draw [arrow] (sign2) -- (abl+1);
        \end{tikzpicture}
    }
\hspace{\fill}
   \subfloat[RBNN \label{fig:RBNN}]
   {%
        \begin{tikzpicture}[node distance=1.2cm]
        \node (Wl+1) [data] {$\mathbf{\overline{W}}_{l}$};
        \node (R) [process_h, right of=Wl+1] {$\mathbf{R}$};
        \node (W'bl+1) [data, right of=R] {$\mathbf{W'}_{l}^b$};
        \node (x) [process, below of=W'bl+1] {$\times$};
        \node (abl) [data, left of=x] {$\mathbf{a}_{l-1}^b$};
        \node (sign2) [process, right of=x] {$\sgn$};
        \node (abl+1) [data, right of=sign2] {$\mathbf{a}_{l}^b$};
        \draw [arrow] (abl) -- (x);
        \draw [arrow] (Wl+1) -- (R);
        \draw [arrow] (R) -- (Wbl+1);
        \draw [arrow] (Wbl+1) -- (x);
        \draw [arrow] (x) -- (sign2);
        \draw [arrow] (sign2) -- (abl+1);
        \end{tikzpicture}
    }\\

\caption{Comparison between the respective MAC operation performed by a real-valued CNN, BNN, and RBNN }
    \label{fig:mac}
\end{figure*}

\begin{table}[!t]
  \begin{center}
    \begin{tabular}{ |p{1.3cm}| p{2cm} | p{4.3cm} |}
 \hline
 \textbf{Layer}  & \textbf{Output Volume}  & \textbf{ Description} \\
 \hline
 input & $1\times2\times1024$ & \\ 
 conv & $32\times2\times1024$ & $32\times(3\times3)$ \\
& & $1\times \text{BN}$ \\
 R-Block A & $32\times2\times1024$ & $2\times[32\times(3\times3)]$ \\
  & & $2\times \text{BN}$, addition \\
 R-Block A & $32\times2\times1024$ & $2\times[32\times(3\times3)]$ \\
 & & $2\times \text{BN}$, addition \\
 R-Block B & $32\times1\times512$ & $1\times[32\times(3\times3)] , 1\times[32\times(1\times1)]$ \\
 & & $1\times[32\times(1\times3)]$\\
 & & $3\times \text{BN}$, addition \\
 R-Block A & $32\times1\times512$ & $2\times[32\times(1\times3)]$ \\
 & & $2\times \text{BN}$, addition \\
 R-Block B & $64\times1\times256$ & $2\times[64\times(1\times3)] , [64\times(1\times1)]$ \\
 & & $3\times \text{BN}$, addition \\
 R-Block A & $64\times1\times256$ & $2\times[64\times(1\times3)]$ \\
 & & $2\times \text{BN}$, addition \\
 R-Block B & $128\times1\times128$ & $2\times[128\times(1\times3)] , [128\times(1\times1)]$
\\
& & $3\times \text{BN}$, addition \\
 R-Block A & $128\times1\times128$ & $2\times[128\times(1\times3)]$ \\
 & & $2\times \text{BN}$, addition \\
 pool & $128\times1\times1$ &  average-pooling $(128)$\\
 BN & $128\times1\times1$ & $1\times \text{BN}$ \\
 linear & $24\times1\times1$ & classification into 24 classes \\ 
 & &  softmax\\
 \hline
\end{tabular}
\caption{Architecture of LResNet}
\label{tab:table2}
\end{center}
\end{table}

\subsection{Binary Neural Network}
Consider a real-valued NN $g_\phi(\cdot)$, where $\phi$ represents the real-valued network parameters. The output of the neural network (NN) is given by $y = g_\phi(x)$, where $x$ is the set of input features to the NN. If $g_\phi(\cdot)$ is a CNN, and let's say that it has $L$ layers, then the parameters of the CNN (filters) are given by $\phi = \{\mathbf{W}_1, \dots, \mathbf{W}_L\}$ where $\mathbf{W}_l \in \mathbb{R}^{c_o \times c_i \times k \times k}$ is the weight matrix for the  $l^{th}$ layer of a two dimensional CNN. Here $c_o$ and $c_i$ represent the input and output channels, and $k$ is the dimension of the filter. The input to the $l^{th}$ layer is $\mathbf{a}_l \in \mathbb{R}^{c_i \times h_{in}^w \times h_{in}^h}$,  where $h_{in}^w$ and $h_{in}^h$ are the width and height of the input, respectively. The output from the $l^{th}$ layer is $\mathbf{a}_{l+1} \in \mathbb{R}^{c_o \times h_{out}^w \times h_{out}^h}$. Here, $h_{in}^w$ and $h_{in}^h$ are the spatial dimensions (width and height) of the input, and $h_{out}^w$ and $h_{out}^h$ are the spatial dimensions (width and height) of the output, respectively. The weights ($\mathbf{W}$) and activations ($\mathbf{a}$) of the BNN are binarized using the sign function before the convolution operation. The binarized parameters corresponding to the $l^{th}$ layer are given by 
\begin{equation}
\begin{aligned}
\mathbf{W}_l^b = \sgn(\mathbf{W}_l), \hspace{5mm}
\mathbf{a}_l^b = \sgn(\mathbf{a}_l),    
\end{aligned}
\end{equation}
where $\mathbf{W}_l$ and $\mathbf{a}_l$ are as discussed previously and $\mathbf{W}_l^b$ and $\mathbf{a}_l^b$ are the binarized version of the parameters, respectively. We can further rewrite the convolution operation as convolution performed with the help of bit-wise operations as follows:

\begin{equation}
\begin{aligned}
\textbf{W}_l \ast \textbf{a}_l \approx \mathbf{W}_l^b \circledast \mathbf{a}_l^b, 
\end{aligned}
\end{equation}

where $\circledast$ is the convolution operation performed with bit-wise operators. Even though the weights are binarized during the forward pass, to be able to perform back-propagation, the latent weights and real-valued gradients are used. The existence of the $\sgn$ function makes it hard to calculate the gradients, and hence it is for the same reason that we use a straight-through estimator to pass the gradient during back-propagation. Suppose if $b = \sgn(r)$, then $\grad_r = \grad_b \mathbf{1}_{|r|\leq1}$ where $\grad_r = \frac{\partial C}{\partial r}$, $\grad_b = \frac{\partial C}{\partial b}$, where C is the cost function of the NN. To maintain a stable update, the real-valued weights are clipped between $\{+1,-1\}$. The seminal work in this area by the authors in \cite{courbariaux2016binarized} was used by us to construct the convolution layers containing binarized weights and activations, converting the floating point operations (FLOPs) to one-bit XNOR and bit count operations. It helped us in changing the setting from real-valued to binary. However, simple binarization results in significant performance degradation, though it leads to significant savings in computation and memory. Hence, in the next section, we discuss how to circumvent this problem of performance degradation.

\subsection{Rotated Binary Neural network}
In this section, we address one of the major shortcomings of a BNN, which is the quantization error that follows due to binarization of the weight vector $\mathbf{w}_l\in \mathbb{R}^{n_l}$ that belongs to the $l^{th}$ layer of the NN where $\mathbf{w}_l$ is the vectorized version of $\mathbf{W}_l$ and $n_l=c_o \cdot c_i \cdot k^2$. The presence of an angular bias between $\mathbf{w}_l^b$ and $\mathbf{w}_l$ could lead to a large quantization error and therefore degrades the performance of the network. To reduce the angular bias, at the beginning of each training epoch, very recently \cite{lin2020rotated} proposed an application of a rotation matrix $\mathbf{R}_l \in \mathbb{R}^{n_{l} \times n_{l}}$ to $\mathbf{w}_l$ such that the angle $\phi_l$ between $(\mathbf{R}_l)^T \mathbf{w}_l$ and its binary vector $\sgn ((\mathbf{R}_l)^T \mathbf{w}_l)$ is minimized. The comparison of this method with that of a traditional BNN and a real CNN is shown in Fig. \ref{fig:mac}. The equation for the angle is formulated as follows:

\begin{figure*}[h!]
    \includegraphics[scale=0.47]{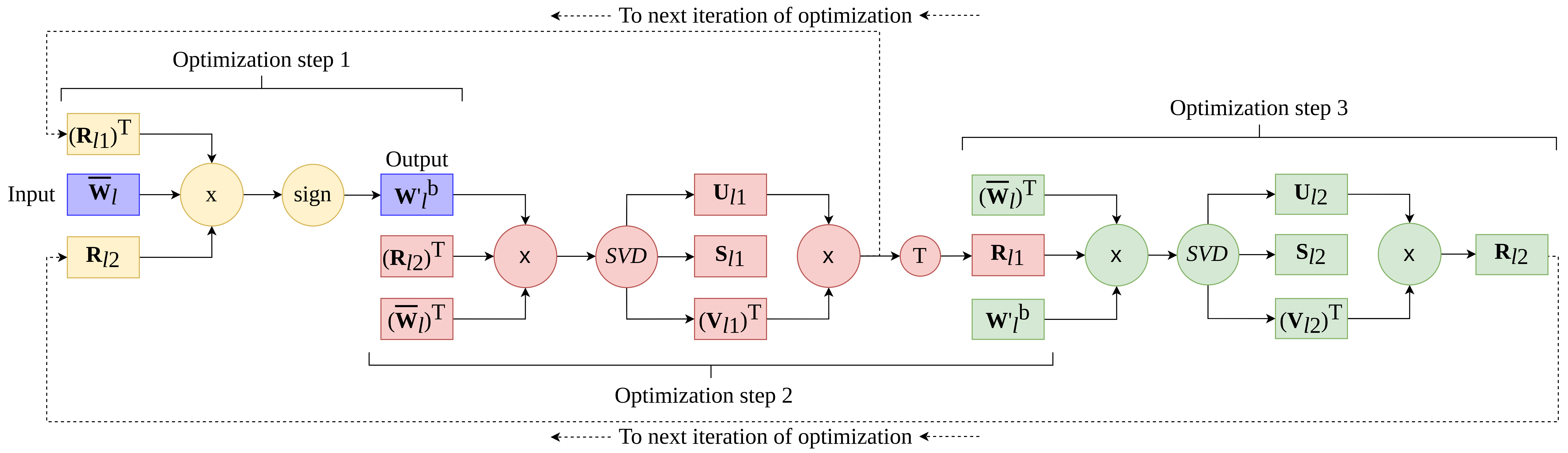}
    \caption{A Flowchart containing the various operations performed as part of the optimization step \textbf{R} mentioned in Fig. \ref{fig:RBNN}.} 
    \label{fig:Optimization}
\end{figure*}

\begin{equation}
\begin{aligned}
\label{eq:cos}
     \cos{(\phi_l)} = \frac{\sgn ((\mathbf{R}_l)^T \mathbf{w}_l)^T ((\mathbf{R}_l)^T \mathbf{w}_l)}{\|\sgn ((\mathbf{R}_l)^T \mathbf{w}_l)\|_2 \|((\mathbf{R}_l)^T \mathbf{w}_l)\|_2},
\end{aligned}
\end{equation}
where $(\mathbf{R}_l)^T\mathbf{R}_l = \mathbf{I}_{n_l}$ is the $n_l$-th order identity matrix. Note, $\|\sgn ((\mathbf{R}_l)^T \mathbf{w}_l)\|_2 = \sqrt{n_l}$ and $\|((\mathbf{R}_l)^T \mathbf{w}_l)\|_2 = \|\mathbf{w}_l\|_2$. Since the training happens at the beginning of each epoch, we can take $\|\mathbf{w}_l\|_2$ to be a constant. Thus, Eq. (\ref{eq:cos}) can be simplified as follows:
\begin{equation}
\begin{aligned}
\label{eq:cosmod}
     \cos{(\phi_l)} &= \eta_l \cdot \sgn ((\mathbf{R}_l)^T \mathbf{w}_l)^T ((\mathbf{R}_l)^T \mathbf{w}_l) \\
      &= \eta_l \cdot \tr (\mathbf{w'}_l^b  (\mathbf{w}_l)^T \mathbf{R}_l),
\end{aligned}
\end{equation}
where $\tr(\cdot)$ is the trace of the input matrix, \begin{equation}\begin{aligned}
    \mathbf{w'}_l^b &= \sgn ((\mathbf{R}_l)^T \mathbf{w}_l) \text{ and}\\ \eta_l &= 1/(\|\sgn ((\mathbf{R}_l)^T \mathbf{w}_l)\|_2 \|((\mathbf{R}_l)^T \mathbf{w}_l)\|_2) \\&= 1/(\sqrt{n_l} \|\mathbf{w}_l\|_2).
\end{aligned}\end{equation}
However, Eq. (\ref{eq:cosmod}) involves a large rotation matrix ($n_l$ can be large), and hence direct optimization of $\mathbf{R}_{l}$ would require massive memory and computation. To reduce that, the authors introduced a scheme using the properties of the Kronecker product where they split the rotation matrix $\mathbf{R}_{l}$ into two smaller rotation matrices $\mathbf{R}_{l1}$ and $\mathbf{R}_{l2}$ that gives
\begin{equation}
\begin{aligned}
    (\mathbf{w}_l)^T (\mathbf{R}_{l1} \otimes \mathbf{R}_{l2}) &= \text{Vec}((\mathbf{R}_{l2})^T (\mathbf{\overline{W}}_l)^T \mathbf{R}_{l1}),
\end{aligned}
\end{equation}
where Vec($\cdot$) vectorizes the input and Vec$(\mathbf{\overline{W}}_l) = \mathbf{w}_l$, $\mathbf{R}_{l1}  \in \mathbb{R}^{n_{l1} \times n_{l1}}\text{, } \mathbf{ R}_{l2}  \in \mathbb{R}^{n_{l2} \times n_{l2}}\text{, } \mathbf{\overline{W}}_l\in\mathbb{R}^{n_{l2} \times n_{l1}} \text{ and } n_{l} = n_{l1} n_{l2}$. Hence, applying a bi-rotation to $\mathbf{\overline{W}}_l$ is equivalent to applying a rotation $\mathbf{R}_{l} = \mathbf{R}_{l1} \otimes \mathbf{R}_{l2}$ to $\mathbf{w}_l$, where $\mathbf{R}_{l} \in \mathbb{R}^{n_{l1}n_{l2} \times n_{l1}n_{l2}}$. Thus, the authors try to find optimal values for $\mathbf{R}_{l1}$ and $\mathbf{R}_{l2}$, which consumes $\mathcal{O}((n_{l1})^2 + (n_{l2})^2)$ space complexity and  $\mathcal{O}((n_{l1})^2n_{l2}+ (n_{l2})^2n_{l1})$ time complexity as compared to $\mathcal{O}((n_l)^2)$ space and time complexity in the absence of bi-rotation, which can make a  huge difference. Further, Eq. (\ref{eq:cosmod}) can be re-written as
\begin{equation}
\begin{aligned}
     \cos{(\phi_l)} &= \eta_l . \tr (\mathbf{w'}_l^b  \text{Vec}((\mathbf{R}_{l2})^T (\mathbf{\overline{W}}_l)^T \mathbf{R}_{l1})) \\ 
     &= \eta_l . \tr (\mathbf{W'}_l^b  (\mathbf{R}_{l2})^T (\mathbf{\overline{W}}_l)^T \mathbf{R}_{l1}),\\
\text{where,\hspace{1cm}} \mathbf{W'}_l^b &= \sgn((\mathbf{R}_{l1})^T\mathbf{\overline{W}}_l\mathbf{R}_{l2}),\\
(\mathbf{R}_{l1})^T \mathbf{R}_{l1} &= \mathbf{I}_{n_{l1}}\\
(\mathbf{R}_{l2})^T \mathbf{R}_{l2} &= \mathbf{I}_{n_{l2}}.
\end{aligned}
\end{equation}

Hence, the optimization objective is given by 
\begin{equation}
\begin{aligned}
\argmax_{\mathbf{W'}_l^b, \mathbf{R}_{l1}, \mathbf{R}_{l2}} &\tr(\mathbf{W'}_l^b  (\mathbf{R}_{l2})^T (\mathbf{\overline{W}}_l)^T \mathbf{R}_{l1})\\
\text{s.t. } \mathbf{W'}_l^b &\in  \{+1,-1\}^{n_{l1}\times n_{l2}} \\
&(\mathbf{R}_{l1})^T\mathbf{R}_{l1} = \mathbf{I}_{n_{l1}} \\
&(\mathbf{R}_{l2})^T\mathbf{R}_{l2} = \mathbf{I}_{n_{l2}}.
\end{aligned}
\end{equation}

Since the above optimization is not a convex problem, the authors proposed an alternating optimization approach, where one variable is updated, keeping the rest two fixed until convergence. We, therefore, have three steps in each cycle:
\begin{enumerate}
    \item The first step is to learn $\mathbf{W'}_l^b$ while fixing $\mathbf{R}_{l1}$ and $\mathbf{R}_{l2}$. Therefore the optimization reduces to
\begin{equation}
\begin{aligned}
    \argmax_{\mathbf{W'}_l^b} \hspace{0.2cm} &\tr (\mathbf{W'}_l^b  (\mathbf{R}_{l2})^T (\mathbf{\overline{W}}_l)^T \mathbf{R}_{l1})\\
    \text{s.t. } \mathbf{W'}_l^b &\in \{+1,-1\}^{n_{l1}\times n_{l2}}\\
    &(\mathbf{R}_{l1})^T\mathbf{R}_{l1} = \mathbf{I}_{n_{l1}} \\
    &(\mathbf{R}_{l2})^T\mathbf{R}_{l2} = \mathbf{I}_{n_{l2}}
\end{aligned}
\end{equation}
which is solved by $\mathbf{W'}_l^b = \sgn((\mathbf{R}_{l1})^T \mathbf{\overline{W}}_l \mathbf{R}_{l2})$.
    
    \item The next step updates $\mathbf{R}_{l1}$ while keeping $\mathbf{W'}_l^b$ and $\mathbf{R}_{l2}$ constant. The corresponding sub-problem is
    \begin{equation}
    \begin{aligned}
    \argmax_{\mathbf{R}_{l1}} \hspace{0.2cm} &\tr (\mathbf{G}_{l1} \mathbf{R}_{l1}) \\
     \text{s.t. } (\mathbf{R}_{l1})^T \mathbf{R}_{l1} &= \mathbf{I}_{n_{l1}},
      \end{aligned}
    \end{equation}
where $\mathbf{G}_{l1} = \mathbf{W'}_l^b  (\mathbf{R}_{l2})^T (\mathbf{\overline{W}}_l)^T$. In order to find the optimal $\mathbf{R}_{l1}$, $\mathbf{G}_{l1}$ is polar-decomposed  using SVD as
\begin{equation}
\begin{aligned}
    \mathbf{G}_{l1} &= \mathbf{U}_{l1} \mathbf{S}_{l1} (\mathbf{V}_{l1})^T
\end{aligned}
\end{equation}
that yields
\begin{equation}
\begin{aligned}
    \mathbf{R}_{l1} = \mathbf{V}_{l1} (\mathbf{U}_{l1})^T.
\end{aligned}
\end{equation}

    \item Similar to the previous steps, the following step updates $\mathbf{R}_{l2}$ while keeping $\mathbf{W'}_l^b$ and $\mathbf{R}_{l1}$ constant. The corresponding sub-problem is
    \begin{equation}
    \begin{aligned}
    \argmax_{\mathbf{R}_{l2}} \hspace{0.2cm} &\tr ((\mathbf{R}_{l2})^T \mathbf{G}_{l2} ) \\
     \text{s.t. } (\mathbf{R}_{l2})^T \mathbf{R}_{l2} &= \mathbf{I}_{n_{l2}}\end{aligned},
    \end{equation}
    where $\mathbf{G}_{l2} = (\mathbf{\overline{W}}_l)^T  \mathbf{R}_{l1} \mathbf{W'}_l^b$. To find the optimal $\mathbf{R}_{l2}$, $\mathbf{G}_{l2}$ is polar-decomposed using SVD as
    \begin{equation}
    \begin{aligned}
    \mathbf{G}_{l2} = \mathbf{U}_{l2} \mathbf{S}_{l2} (\mathbf{V}_{l2})^T
    \end{aligned}
    \end{equation}
    that yields 
    \begin{equation}
    \begin{aligned}
    \mathbf{R}_{l2} = \mathbf{U}_{l2} (\mathbf{V}_{l2})^T.
    \end{aligned}
    \end{equation}
\end{enumerate}

The above-mentioned optimization steps are performed iteratively, as shown in Fig. \ref{fig:Optimization}. As discussed in \cite{lin2020rotated}, $\mathbf{W'}_l^b$, $\mathbf{R}_{l1}$, and $\mathbf{R}_{l2}$ converge within a maximum of three cycles. However, the above optimization could get caught in a local optimum. Hence, the adjustable rotated weight vector scheme was proposed to reduce the angular bias after the bi-rotation step \cite{lin2020rotated}. Instead of using the rotated weights as they are,
\begin{equation}
\begin{aligned}
\Tilde{\mathbf{w}_l} = (\mathbf{R}_l)^T \mathbf{w}_l,
\end{aligned}
\end{equation}
they propose the usage of the following update equation 
\begin{equation}
\begin{aligned}
\label{eq:weightupdate}
\Tilde{\mathbf{w}_l}    = \mathbf{w}_l + ((\mathbf{R}_l)^T \mathbf{w}_l - \mathbf{w}_l) \cdot \alpha_l,\\
\text{where } \alpha_l = |\sin{(\beta_l)}| \in [0,1] \hspace{0.15cm} \text{and} \hspace{0.15cm} \beta_l \in \mathbb{R}.
\end{aligned}
\end{equation}
During training the RBLResNet, at the beginning of every training epoch, the rotation matrices, $\mathbf{R}_{l1}$ and $\mathbf{R}_{l2}$, are learned for a fixed $\mathbf{w}_l$. At the training phase, with the fixed $\mathbf{R}_{l1}$ and $\mathbf{R}_{l2}$, the NN takes the sign of parameter $\tilde{\mathbf{w}}_l$ for the forward pass and the parameters $\mathbf{w}_l$ and $\beta_l$ are updated during back-propagation. Since $\beta_l$ is also a trainable parameter, it enables the network to learn a suitable value of $\alpha_l$ that further optimizes the application of rotation. In RBLResNet, the rotated binarization is applied to the parameters of all the convolutional layers except those next to the input and output layers. The first convolution layer is left to be a real layer to extract features from the input more efficiently, and the last linear layer is also real. The performance and saving in memory and computation of RBLResNet are discussed in the next subsection.

\subsection{Savings in Computation} \label{subsec:savings}

The work in \cite{9505671} has approximated the number of multiplication and addition operations performed as part of a single convolutional layer for a 1DCNN during run-time. Here, we extend the same for a 2DCNN. The convolution between real-valued $\mathbf{W}_{l} \in \mathbb{R}^{c_o \times c_i \times k \times k}$ and $\mathbf{a}_{l} \in \mathbb{R}^{c_i \times h_{in}^w \times h_{in}^h}$ layer results in an output $\mathbf{a}_{l+1} \in \mathbb{R}^{c_o \times h_{out}^w \times h_{out}^h}$. The total number of multiplication for the $l^{t h}$ layer is $c_i \times k^2 \times h^w_{out}\times h^h_{out} \times c_o$ and the total number of addition for the $l^{t h}$ layer is $\left(c_{i}-1\right) \times(k^2-1) \times h^w_{out} \times h^h_{out} \times c_{o}$. The total count of FLOPs for the $l^{t h}$ layer of a real-valued 2DCNN is the summation of the number of multiplication and addition that is roughly twice the number of multiplication given by $2 \times c_i \times k^2 \times h^w_{out}\times h^h_{out} \times c_o$. 

The primary motivation of the proposed system is to save on computational complexity and memory. We achieve that by using BNNs and RBNNs with better performance using the latter. BNN and RBNN convert weights and activations into binary $\{+1,-1\}$, making it possible to carry out convolution operations using the efficient XNOR and bit-count logic instead of FLOPs. A single FLOP count operation needs one $64$-bit register for the $64$-bit floating point operation. The $64$ single-bit XNOR-count operations can be computed using the same $64$-bit register, which makes the system almost $64$ times faster than the real networks. For a $64$-bit system, the same number of real-valued parameters take $64$ times more memory than binary parameters in BNNs and RBNNs. These advantages make the system a perfect fit for edge devices that are typically limited in memory and power. Although RBLResNet has minimal memory and computation, the performance gap compared to real networks is still significant. In the following subsection, we try to close the gap further by using two different ensemble techniques.

\subsection{Multilevel classification and Ensemble of RBNNs}
In this section, we propose two different methods, multilevel classification and bagging, to further improve the performance of the RBLResNet.

\subsubsection{Multilevel Classification}
As AMC is a complex classification problem involving many modulation classes, we propose to use MC that divides the problem into multiple simpler sub-problems and solves them in different levels. MC is a form of the well-studied hierarchical classification problem \cite{silla2011survey}. Let the modulation classification task have $K$ classes and $M$ sub-problems; each of the $M$ sub-problems solves the classification with $K/M$ modulation classes. 

Thus, the problem is now split into two levels: the first classifies the input data into different clusters, followed by the second level, which identifies the target modulation scheme within each cluster. By doing so, we deal with a single $M$-class problem and $M$ such $K/M$-class problems. This results in a total of $M+1$ sub-problems, each of which is solved with the help of a unique RBLResNet trained separately for every sub-problem. This way, if required, each sub-problem can be solved by different networks of varying complexity based on the difficulty of the problem. 

The network trained using MC, where the networks corresponding to all the sub-problems are identical, is called RBLResNet-MC. We also introduce RBLResNet-MCK, in which the network corresponding to one of the sub-problems is replaced with an RBLResNet with more filters. This technique is introduced to improve the performance in case one of the sub-problems has a lower classification accuracy.

\subsubsection{Bagging}

Ensemble methods reduce bias and variance by combining multiple models, known as base learners, to improve accuracy and reduce overfitting. It has been established that ensemble methods such as bagging, boosting, and stacking lead to a boost in accuracy \cite{ganaie2022ensemble}. In the context of AMC, we propose an ensemble of multiple RBLResNets using bagging.

Bagging is a model averaging method known to improve stability and accuracy and reduce variance \cite{breiman1996bagging}. Let the outputs of the penultimate layers (the linear layers in Fig. \ref{fig:Architecture}) of $B$ RBLResNets be denoted as  {$\hat{u}^1, \cdots,\hat{u}^B$}. To perform bagging, we compute the weighted average of these outputs as $\hat{u}=\Sigma_{b=1}^{B}w^b\hat{u}^b$ such that $\Sigma_{b=1}^{B}w^b=1$. This averaged output is then passed through a softmax function and used for classification. Although bagging $B$ such networks would improve the accuracy, the memory requirements and computational complexity also increase by $B$. However, the computation time remains the same as a single network, as the $B$ networks may be implemented in parallel.

\subsection{Adversarial Robustness}
Until now, we have proposed methods to bridge the performance gap between real and rotated binarized versions. We now discuss the adversarial robustness of our proposed methods. It has been found that the addition of malicious perturbations to the input of a DNN can cause it to misclassify data \cite{szegedy2013intriguing}. These malicious perturbations are referred to as adversarial attacks. Being robust to these attacks is essential, especially for critical applications like AMC.

The authors in \cite{galloway2018attacking} first explored robustness to adversarial attacks in the case of binarized networks. The authors attribute the adversarial robustness of BNNs to the following: 
\begin{itemize}
\item As the weights and activations are restricted to $\pm1$, the network is implicitly regularized. \item BNNs are harder to train; hence, they are harder to attack through iterative attacks. \item BNNs exhibit a higher degree of nonlinearity.
\end{itemize}
All these properties also hold for our proposed rotated binarized network, making them robust to adversarial attacks. We demonstrate the adversarial robustness of our proposed RBLResNet through experimental evaluation in Section \ref{sec:radioML}. 
 
We now explore further improving the adversarial robustness while constructing an ensemble network. It has been found that the local Lipschitz constant (defined below) of a network is inversely proportional to its adversarial robustness \cite{yang2020closer}.

\noindent \textit{Definition:} 
A function \(f:\mathbb{R}^m \rightarrow \mathbb{R}^n\)  is said to be \(L_f\)- locally Lipschitz over $\mathcal{X} \subseteq \mathbb{R}^m$ if  \(\forall\) \(\bm{x_1},\bm{x_2} \in \mathcal{X}\),
    \begin{equation*} 
    \label{eqn:LipDefn}
    \|f(\bm{x_1}) -  f(\bm{x_2})\| \leq L_ f \|\bm{x_1-x_2} \|,
    \end{equation*}
where \(L_f\) is the local Lipschitz constant (LC).
Exploiting the relationship between the local LC and adversarial robustness, \cite{tholeti2022robust} proposes constructing an ensemble system with better adversarial robustness. To further improve  the adversarial robustness of our proposed ensemble method, we employ \cite{tholeti2022robust} as described below.

Recall that bagging is performed by evaluating the weighted average of the outputs of the constituent networks, also known as base learners. The first step towards creating a robust bagged network is to calculate the local Lipschitz constant\footnote{We discuss the method for empirically estimating the local Lipschitz constant in Sec. \ref{sec:radioML}.} for each of the base learners, which are RBLResNets for our application. The base learners are then weighed such that their weights are inversely proportional to the corresponding local Lipschitz constants\footnote{ Note: Even though the architectures are the same for all base learners, they are trained independently. As a result, their local Lipschitz constants are not exactly the same. }. Given that the output of a base learner $b$ is given by $\hat{u}^b$, the output of the bagged ensemble is given by 
\[\hat{u}=\Sigma_{b=1}^{B}w^b\hat{u}^b \quad \text{s.t.  } w^b \propto \dfrac{1}{L_b} \text{ and } \Sigma_{b=1}^{B}w^b=1.\]
To summarize, a base learner with a lower local Lipschitz constant and higher adversarial robustness will be weighed more. In the following section, we provide experimental results for all the proposed models and compare them with state-of-the-art techniques.

\begin{table*}[h!]
  \begin{center}
    \begin{tabular}{|p{2.7cm}|p{3.2cm} |p{4.5cm}| p{1.4cm}| p{1.4cm}|p{2.0cm}|}
 \hline
 \textbf{Setting} & \textbf{Model} &  \textbf{Computation {(in equivalent FLOPs)\textbf{*}}}  &
 \textbf{Parameters} &\textbf{Memory Req (MB)} & \textbf{Clean Accuracy $\%$ at 10dB} \\
 \hline

\multirow{9}{8em}{Real valued}& RMLResNet \cite{o2018over} & $\simeq 1.3e8$ FLOPs & $2.37e5$&$1.9$& $91.47$\\

& Lightweight \cite{kim2020lightweight} & $\simeq 8.9e7$ FLOPs & $5.00e4$&$0.4$ & $91.48$\\

& MCNet ($m=10$) \cite{8963964}  & $\simeq 1.9e7$ FLOPs  &$2.20e5$&$1.76$& $92.25$ \\

& MLDNN\cite{181} & $\simeq 1.6e10$ FLOPs  & $8.99e5$&$7.2$ & $92.4$\\


& MLResNet \cite{1810} & ${\simeq 8e9}$ FLOPs & ${1.15e5}$&${0.92}$ & ${94.5}$\\

& InvoResNet \cite{zhang2021automatic} & $\simeq 1.7e10$ FLOPs& $2.20e5$&$1.76$ &$94.6$ \\

& LResNet (Ours)& $\simeq 2.4e8 $ FLOPs  &$2.88e5$&$2.2$ &$95.79$ \\


& ASN\cite{182} & ${\simeq 8e8}$ FLOPs & ${5.4e7}$& ${432}$ & ${98.2}$\\

\hline
BNN & BLResNet &$\simeq 3.8e6 $ FLOPs &$2.88e5$&$0.03$ &$42.75$ \\

BNN with 2 real layers  & BLResNet2R & $\simeq 4.9e6 $ FLOPs & $2.88e5$ & $0.06$ &$54.76$ \\
\hline

& RMLResNet & $\simeq 2.9e6$ FLOPs & $2.37e5$&$0.03$ &$59.53$\\

& MCNet ($m=10$)  & $\simeq 2.8e6 $ FLOPs &$2.20e5$&$0.03$& $62.2$ \\

RBNN & RBLResNet (Ours) & $\simeq 4.9e6 $ FLOPs  & $2.88e5$ & $ 0.06$ &$ 86.95$ \\

& RBLResNet-Bag2 (Ours)& $\simeq 2 \times (4.9e6 $ FLOPs)** & $5.76e5$&$0.13$ &$ 88.42$ \\

& RBLResNet-Bag4 (Ours)& $\simeq 4 \times (4.9e6 $ FLOPs)** & $1.15e6$&$0.25$ &$ 89.70$ \\

& RBLResNet-MC (Ours)& $\simeq 2 \times (4.9e6 $ FLOPs) & $1.15e6$&$0.25$ &$ 91.95$ \\
\rowcolor{yellow}
& RBLResNet-MCK (Ours)& $\simeq  1.4e7 $ FLOPs & $2.01e6$&$0.37$ &$ 93.39$ \\
\hline
\end{tabular}
\caption{Savings vs. performances of different networks on the RML2018.01a dataset. The computation shown in the table accounts only for the methods' run-time and not training time since the resource-constrained edge network is not expected to perform training. We have highlighted our proposed method with the best performance in the low-complexity range. \textbf{*}Note, a total of $64$ XNOR count operations can be performed at the same time as $1$ FLOP
in systems with $64$ bit registers, the computations of the BNN and RBNN setting containing XNOR counts have been converted to equivalent no. of FLOPs.
\textbf{**}Even though the computation for the ensemble increases by the factor of the number of networks in the ensemble, the computation time remains the same as that of a single network due to parallel processing capability. }
\label{tab:allmethods}
\end{center}
\end{table*}

\section{Experimental evaluation} \label{sec:radioML}
\subsection{Experimental Setup}
In this section, we provide numerical results to investigate the performance of the proposed AMC methods. Some existing ML-based AMC solutions have been benchmarked using an older dataset version from Deepsig\footnote{https://www.deepsig.ai/datasets} with only $11$ modulation schemes \cite{rajendran2018deep,9194036}. The industry-standard data set, and its following updates for modulation classification in radio, are given by \cite{o2016convolutional}, \cite{o2016unsupervised}. In \cite{o2018over}, an updated version of the RML$2018.01$a (R: Radio; ML: Machine Learning) dataset was released. 

 The latest release is one of the most challenging datasets of modulation classification. It includes higher-order modulation schemes (QAM$256$ and APSK$256$) used in the real world in high-SNR low-fading channel environments. The dataset comprises simulated channel effects created artificially (such as carrier frequency shift, variation in symbol rate, signal delay, and thermal noise) and real-world measurements. The synthetic data was produced using software-defined radio programmed with GNU radio \cite{10.5555/993247.993251}. 
The dataset has $24$ modulations schemes, namely OOK, 4ASK, 8ASK, BPSK, QPSK, 8PSK, 16PSK, 32PSK, 16APSK, 32APSK, 64APSK, 128APSK, 16QAM, 32QAM, 64QAM, 128QAM, 256QAM, AM-SSB-WC, AM-SSB-SC, AM-DSB-WC,AM-DSB-SC, FM, GMSK, OQPSK, with $26$ evenly spaced bins in signal-to-noise ratio (SNR), ranging from $-20$ to $30$dB. The set comprises $2,555,904$ I/Q (in-phase/quadrature) signals, each of length $1024$ (array shape is $2 \times 1024$).

All the experiments were performed using a $3.00$ GHz CPU, $64$ GB RAM, and an NVIDIA Geforce RTX 2080Ti GPU. RBNN is trained end-to-end with random initial weights for $500$ epochs using the stochastic gradient descent optimizer with a momentum of $0.9$. The mini-batch size for each iteration is set to $256$, and the learning rate is initialized at $0.01$. The experiments use the cosine-annealing learning rate scheduler, where the learning rate is often restarted to simulate a warm restart. For LResNet, we use a multi-step learning rate scheduler with a decay factor of $10$ and trained for $200$ epochs. Out of the complete dataset, $75\%$ of the samples are used for training and the rest for testing. We use the categorical cross-entropy loss function.

We also evaluate the performance of our proposed models under adversarial attacks. For the attacks, we have considered gradient-based white box attacks FGSM and PGD. Classifiers frequently misclassify negative SNRs because of poor signal quality in real-world situations; hence, we have only considered the positive SNRs (0 to 30dB) for evaluating adversarial attacks. We also need to estimate the local Lipschitz constant for each base learner to create the ensemble bagging network. The empirical evaluation of the local LC is described below.
\subsubsection{Measuring Local Lipschitz Constant}
 To empirically evaluate the local LC of a network, we consider a dataset $(\mathbf{x_i},y_i)$ where $i=1,\cdots,n$. The local Lipschitz constant is evaluated for a function $f$, which is represented by the following expression (as per \cite{yang2020closer})
\begin{equation*}
    L_f=\dfrac{1}{n}\sum_{i=1}^{n} \underset{\mathbf{x_{i}'}\in B_{\infty(\mathbf{x_{i}},\mu)}}{\mathrm{max}}\dfrac{||f(\mathbf{x_i})-f(\mathbf{x'_i})||_1}{||\mathbf{x_i-x'_i}||_\infty}.
\end{equation*}
Here, the perturbed input $\mathbf{x'_i}$ is taken from an infinity norm-ball $B$ around $\mathbf{x_i}$ with a perturbation radius of $\mu$. Note that $\mu$ is a hyperparameter; for our experiments, we use $\mu =0.03$. The operation $\max$ in the above expression is empirically solved by adopting an iterative gradient approach by moving in the gradient direction given by $\nabla_{\mathbf{x'_i}}\dfrac{||f(\mathbf{x_i})-f(\mathbf{x'_i})||_1}{||\mathbf{x_i-x'_i}||_\infty}$. This iterative approach employs a step size of $\mu/5$ and $10$ steps.

\subsection{ Experimental Results}
\subsubsection{Performance and savings}
\begin{figure}[ht]
    \centering
    \includegraphics[scale=0.65]{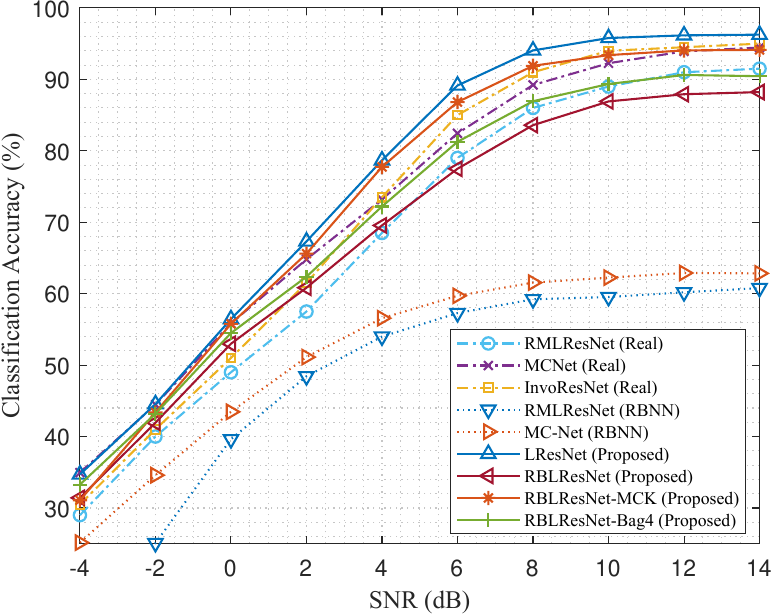}
    \caption{Accuracy vs. SNR for different architectures}
    \label{fig:ACC vs snr}
\end{figure}

On RML2018.01a, the proposed RBLResNet and its ensemble versions are compared with several SOTA architectures. In the case of MCNet and RMLResNet, we also simulate the RBNN versions of the networks for a more thorough comparison\footnote{Codes for all the experiments are available at \url{https://github.com/deepsy1998/RBLResNet}.}. We list the FLOPs, memory requirements, and clean accuracy at 10dB in Table \ref{tab:allmethods} and the performance of different architectures across a range of SNRs in Fig. \ref{fig:ACC vs snr}.

As can be seen from Table \ref{tab:allmethods}, the proposed real-valued LResNet architecture outperforms several of the SOTA architectures. For example, where MCNet with $10$ M-blocks and RMLResNet have an accuracy of $92.25\%$ and $91.47\%$, respectively, the proposed LResNet has an accuracy of $\mathbf{95.79\%}$. The recently proposed InvoResNet has $94.6\%$ accuracy and costs $70.83$ times more computational complexity than the proposed real architecture. Compared to existing SOTA architectures, the proposed LResNet has a larger number of parameters, which makes it suitable for binarization. However, the deployment of LResNet is suitable only when the devices have enough memory and computing resources. We have also compared the proposed LResNet with three other very recent works (i) ASN\cite{182}, (ii) MLResNet\cite{1810} and (iii) MLDNN \cite{181} in Table \ref{tab:allmethods}. These architectures are unsuitable for deployment at the edge since they have huge memory requirements or computational complexity.

We now discuss the performance of the binarized versions of the aforementioned real networks. Although the binarized version (BLResNet) provided a classification accuracy of $42.75\%$ at $10$dB, the binarized version of MCNet and the RMLResNet architectures failed to learn any meaningful features and performed close to random; thus, those results are not included in Table \ref{tab:allmethods}. This suggests that binarizing any real network blindly may not preserve accuracy. It is also supported by \cite{rajendran2018deep}, which shows that binarizing LSTM or traditional CNN does not yield good results when used for AMC. We also consider a hybrid architecture named BLResNet2R, where the layers adjacent to the input and output layers are real; this is similar to the RBNN architecture and helps to transfer more information from the input layer to the network. It is observed that it improves the performance with $54.76\%$ accuracy at $10$dB.

The proposed RBLResNet architecture achieves an accuracy of $86.95\%$ with a significant improvement of $24.75\%$ and $27.42\%$ over the RBNN versions of MCNet and RMLResNet at $10$ dB SNR while saving on both memory and computational complexity. The RBLResNet has a memory requirement of $0.06$ MB, which is lower than any other real network. Lightweight \cite{kim2020lightweight} is one of the few existing works in which a real network achieves an accuracy of $91.48\%$ at $10$dB with less computation and memory requirement. But while taking a rotated binarization of Lightweight's architecture, it did not learn anything meaningful. The proposed method RBLResNet has $18$ times lesser computation and $6.7$ times lesser memory requirement than Lightweight. 

Further, we implemented ensemble methods, namely bagging and multilevel classification, to bridge the gap between RBLResNet and the real networks. Recall that the bagged network that uses two and four RBLResNets are called RBLResNet-Bag2 and RBLResNet-Bag4, respectively. The $10$dB accuracy for proposed RBLResNet-Bag2 and RBLResNet-Bag4 are $88.42\%$ and $89.70\%$, respectively.

We further use multilevel classification to divide the 24 modulation schemes into three clusters with eight modulation classes each. This reduces the problem into one subproblem of dividing the 24 classes into three equal clusters and three subproblems of classifying those eight classes from a cluster. Our RBLResNet is trained specifically for each subproblem, giving better accuracy than bagging. Multilevel classification helps us gain a massive $5\%$ over the standard RBLResNet at $10$dB, with an accuracy of $91.95\%$. While implementing the multilevel classification, we noted that two out of the three clusters provided better accuracy. To address this, we propose RBLResNet-MCK, employing an architecture with more filters for that cluster with weaker performance. This technique yields an accuracy of $93.39\%$ better than the vanilla multilevel classification. The proposed RBLResNet-MCK performs better than real-valued Lightweight, saving $6.36$ times in computations.

In Fig. \ref{fig:ACC vs snr}, we see the performance of different architectures over a range of SNRs. The proposed LResNet outperforms all the other methods, closely followed by RBLResNet-MCK, which performs better than all other SOTA networks for almost the entire range. At lower SNRs, RBLResNet-Bag4 has comparable accuracy with RBLResNet-MCK and performs better than other real architecture. The RBNN versions of MCNet and RMLResNet perform poorly for the entire range, and their performance falls sharply at lower SNRs, which further proves that simply binarizing any network does not guarantee good performance.

\begin{figure}[!t]
    \centering
    \includegraphics[scale=0.65]{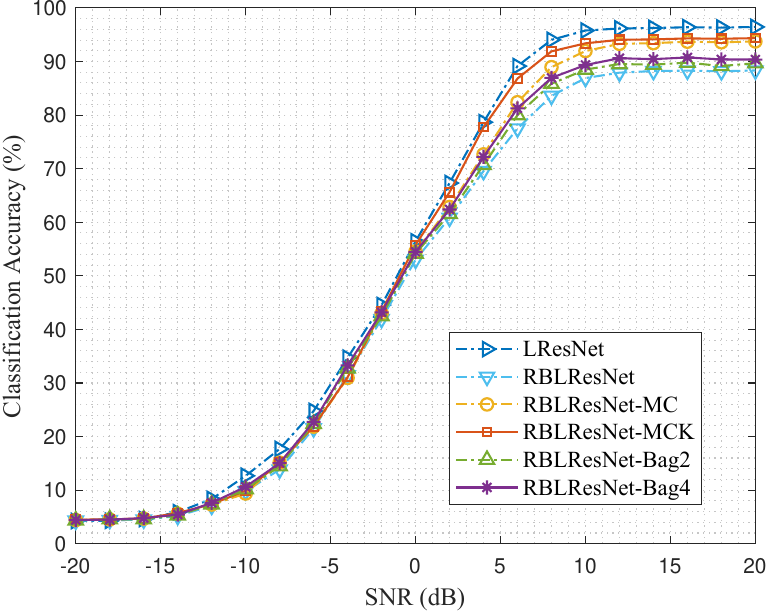}
    \caption{Improvement in accuracy by taking ensemble}
    \label{fig:ACC vs snr2}
\end{figure}

We acknowledge that ensemble methods increase performance with increasing complexity. However, Multilevel classification performs better than bagging with lesser complexity compared to the RBLResNet-Bag 4 network. The performance of ensemble networks is shown in Fig. \ref{fig:ACC vs snr2} for a range of SNRs. It can be seen that bagged ensemble methods perform better than single RBLResNet over the entire range of SNRs. Further, RBLResNet-MCK almost catches up with LResNet over the SNR range (-2 to 6dB).

\begin{figure}[ht]
    \centering
    \includegraphics[scale=0.62]{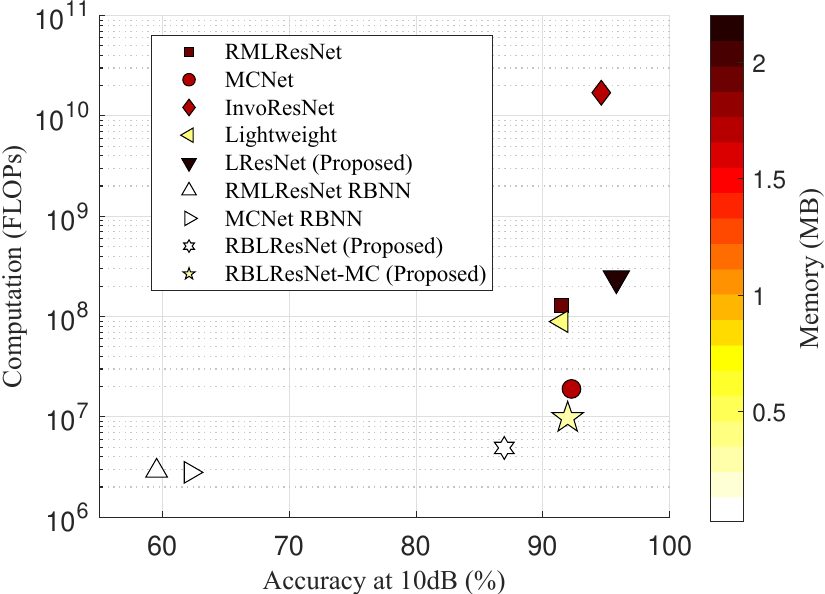}
    \caption{Computation vs. Accuracy at 10dB vs. Memory}
    \label{fig:Performance plot}
\end{figure}

In Fig. \ref{fig:Performance plot}, we have compared the computation, accuracy, and memory of different architectures. The methods in the bottom right of the scatter plot have maximum accuracy and require the least computing power. The color-map shows the memory requirement. It clearly shows that the proposed RBLResNet and its ensembled versions have the lowest computation and memory requirements while achieving accuracies better than some SOTA architectures. RBLResNet-MCK is at the bottom rightmost part of the plot, which shows its superiority over other architectures while considering memory, computational complexity, and accuracy.

\subsubsection{Adversarial Robustness}
Besides being memory and computation-efficient, RBLResNet is more adversarially robust than all SOTA architectures and the real version LResNet. Table \ref{tab:advacc} gives the adversarial accuracy of different architectures under attack. We use adversarial samples generated with FGSM and PGD attacks, which are well-known attacks in machine learning literature. LResNet has an adversarial accuracy of $67.8\%$ for FGSM attacks and $77.7\%$ for PGD attacks. The proposed RBLResNet has a better adversarial accuracy of $73.87\%$ under the FGSM attack and $78.69\%$ under the PGD attack. If we study the local Lipschitz constant for both LResNet and RBLResNet, we see that the local Lipschitz constant for LResNet is estimated at $7.47$, whereas for RBLResNet, it is estimated at $3.10$. This indicates that RBLResNet with a lower local Lipschitz constant is more robust than its real counterpart. Further, we improve the robustness against attacks using ensemble methods.

Typically, in machine learning problems, the adversarial samples are generated from each of the base learners of an ensemble system. These adversarial samples are fed to the corresponding base learners that they are generated from. For our AMC application, we are restricted to a single receiver system. This means the same copy of the received signal is fed to all base learners in an ensemble system. Therefore, to evaluate the adversarial robustness of the system, we generate the adversarial samples from only one of the base learners (chosen at random). The generated adversarial samples are then fed as input to all the learners in the ensemble system, and adversarial accuracy is measured.

\begin{figure}[t]
    \centering
    \includegraphics[scale=0.65]{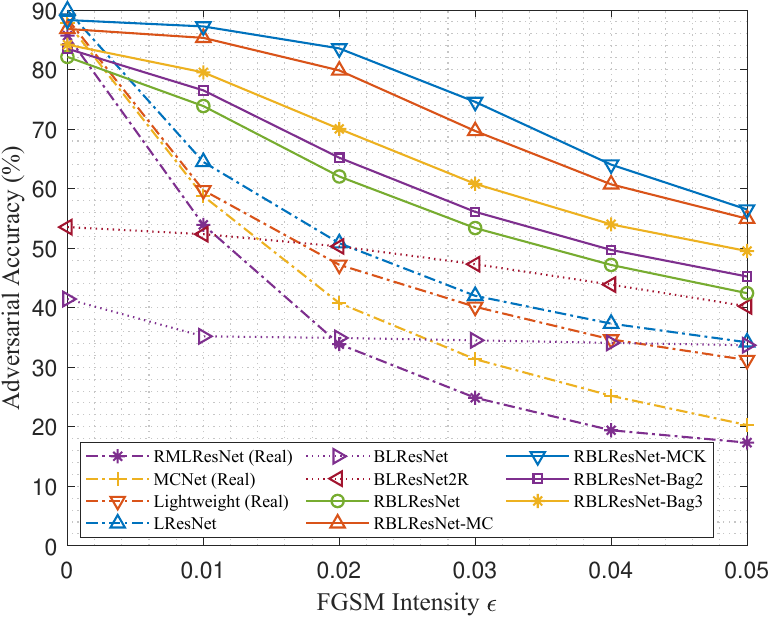}
    \caption{Performance of different architectures under FGSM attack.}
    \label{fig:Adversarial plot}
\end{figure}

\begin{table}[!t]
  \begin{center}
    \begin{tabular}{|p{1.3cm}|p{3.0cm} |p{1.1cm}|p{1.0cm}|}
 \hline
 \textbf{Setting} & \textbf{Model} & \textbf{FGSM Attack} & \textbf{PGD Attack} \\
 \hline

\multirow{3}{8em}{Real valued}& RMLResNet \cite{o2018over} & $53.91$ & $68.32$\\

& Lightweight \cite{kim2020lightweight} & $59.72$ & $71.85$\\

& MCNet ($m=10$) \cite{8963964}  & $58.74$ & $8.15$ \\

& LResNet (Ours) &$67.80$ & $77.77$\\

\hline

BNN  & BLResNet &$35.19$ & $41.26$\\

     & BLResNet2R &$52.35$ & $53.07$\\

\hline

     & RMLResNet  &$54.68$ & $56.50$\\

     & MCNet ($m=10$) & $56.41$ & $58.56$\\

RBNN & RBLResNet (Ours) &$ 73.87$ & $78.69$ \\

     & RBLResNet-Bag2 (Ours) &$76.50$ & $80.70$\\

     & RBLResNet-Bag3 (Ours)  &$ 79.53$ & $82.30$\\
     
     & RBLResNet-MC (Ours) &$ 85.32$ & $86.40$ \\

	 \rowcolor{yellow}
	 & RBLResNet-MCK (Ours) &$87.25$ & $87.95$ \\
\hline
\end{tabular}
\caption{Adversarial robustness of different networks on the RML2018.01a dataset. The adversarial accuracy is calculated over all positive SNRs under the FGSM attack ($\epsilon = 0.01$) and PGD Attack ($ \eta = 0.005, \xi = 10$). We have highlighted the method with the highest adversarial accuracy.}
\label{tab:advacc}
\end{center}
\end{table}

With RBLResNets as base learners, bagging two and three such RBLResNets give an improvement of $2.64\%$ and $5.66\%$, respectively, in adversarial accuracy under FGSM attack over a single RBLResNet. Under the PGD attack, the performance improvement is $2.01\%$ and $3.61\%$ for RBLResNet-Bag2 and  RBLResNet-Bag3, respectively. To achieve better adversarial robustness for RBLResNet-Bag2 and RBLResNet-Bag3, instead of giving all models equal weightage, we have weighted them inversely proportional to their local Lipschitz constants. In addition to lowering the local Lipschitz constant of the ensemble, this provides an added advantage that the attacker does not know how the models are weighed; hence it would be even more challenging to attack the system. Lipschitz weighted ensemble system takes the same memory and computational complexity as conventional bagging with equal weightage but provides an added advantage of more robustness.

Further, RBLResNet-MC and its superior version RBLResNet-MCK seem to have better adversarial accuracy owing to their more substantial generalization power since the classification problem has been broken down into several sub-problems. In the case of Multilevel classification, the adversarial samples have been generated from the learner in the first level. RBLResNet-MCK stands best among the compared methods, with an accuracy of $\mathbf{87.25\%}$  and $\mathbf{87.95\%}$ under FGSM and PGD attacks.

Fig. \ref{fig:Adversarial plot} shows the drop in accuracy for different architectures under increasing attack intensity. The existing real-valued architectures have good clean accuracy, but their performances drop drastically under attack. The drop in accuracy is not significant for the binarized versions. The proposed RBLResNet and its ensembled methods have a good balance between clean and adversarial accuracy. RBLResNet-MCK has the highest adversarial accuracy across all attack intensities in Fig. \ref{fig:Adversarial plot}.

    \section{Conclusion}
We first proposed the LResNet architecture suitable for binarization; it reduces the complexity to make the resulting BLResNet deployable at the edge. To close the performance gap between LResNet and BLResNet, we proposed the rotated binarization for our architecture, namely RBLResNet. This architecture closes the performance gap to achieve an accuracy nearly as good as the existing SOTA architectures. The memory requirement and the computations of RBLResNet are $64$ times lesser than that of LResNet. To further improve the accuracy, we proposed two ensemble techniques bagging and Multilevel classification (MC) using the proposed RBLResNet as its constituents. The proposed MC method, RBLResNet-MCK, produces an accuracy of $93.39\%$, comparable to the InvoResNet\cite{zhang2021automatic} architecture with a $4.75$ times lower memory and $1214$ times lesser computing power which makes it an excellent choice for AMC at edge devices. Further, we showed that RBLResNet is adversarially more robust than real SOTA networks. The proposed RBLResNet has an adversarial accuracy of $73.87\%$, whereas SOTA architectures like Lightweight\cite{kim2020lightweight} have an accuracy of only $59.72\%$. Moreover, RBLResNet-MCK achieved a high adversarial accuracy of $87.25\%$. With a proper balance of clean and adversarial accuracy, RBLResNet-MCK is an ideal choice for achieving a robust resource-constrained AMC architecture at the edge network.
        
 	\bibliographystyle{IEEEtran}
    \bibliography{library.bib}
\end{document}